%
\let\useblackboard=\iftrue
%
%
\newfam\black
\input harvmac.tex
\def\Title#1#2{\rightline{#1}
\ifx\answ\bigans\nopagenumbers\pageno0\vskip1in%
\baselineskip 15pt plus 1pt minus 1pt
\else
\def\listrefs{\footatend\vskip 1in\immediate\closeout\rfile\writestoppt
\baselineskip=14pt\centerline{{\bf References}}\bigskip{\frenchspacing%
\parindent=20pt\escapechar=` \input
refs.tmp\vfill\eject}\nonfrenchspacing}
\pageno1\vskip.8in\fi \centerline{\titlefont #2}\vskip .5in}

\ifx\answ\bigans\def\tcbreak#1{}\else\def\tcbreak#1{\cr&{#1}}\fi
\useblackboard
\message{If you do not have msbm (blackboard bold) fonts,}
\message{change the option at the top of the tex file.}
\font\blackboard=msbm10 scaled \magstep1
\font\blackboards=msbm7
\font\blackboardss=msbm5
\textfont\black=\blackboard
\scriptfont\black=\blackboards
\scriptscriptfont\black=\blackboardss

\else

\fi
\def\yboxit#1#2{\vbox{\hrule height #1 \hbox{\vrule width #1
\vbox{#2}\vrule width #1 }\hrule height #1 }}
\def\fillbox#1{\hbox to #1{\vbox to #1{\vfil}\hfil}}
\def\ybox{{\lower 1.3pt \yboxit{0.4pt}{\fillbox{8pt}}\hskip-0.2pt}}
\def\np#1#2#3{Nucl. Phys. {\bf B#1} (#2) #3}
\def\pl#1#2#3{Phys. Lett. {\bf #1B} (#2) #3}

\def\physrev#1#2#3{Phys. Rev. {\bf D#1} (#2) #3}

\def\comments#1{}

\def\CM{{\cal M}}

\def\CW{{\cal W}}

\def\II{\relax{I\kern-.07em I}}

\def\IZ{\relax\ifmmode\mathchoice
{\hbox{\cmss Z\kern-.4em Z}}{\hbox{\cmss Z\kern-.4em Z}}
{\lower.9pt\hbox{\cmsss Z\kern-.4em Z}}
{\lower1.2pt\hbox{\cmsss Z\kern-.4em Z}}\else{\cmss Z\kern-.4em
Z}\fi}
\def\IB{\relax{\rm I\kern-.18em B}}
\def\IC{{\relax\hbox{$\inbar\kern-.3em{\rm C}$}}}
\def\ID{\relax{\rm I\kern-.18em D}}
\def\IE{\relax{\rm I\kern-.18em E}}
\def\IF{\relax{\rm I\kern-.18em F}}
\def\IG{\relax\hbox{$\inbar\kern-.3em{\rm G}$}}
\def\IGa{\relax\hbox{${\rm I}\kern-.18em\Gamma$}}
\def\IH{\relax{\rm I\kern-.18em H}}
\def\II{\relax{\rm I\kern-.18em I}}
\def\IK{\relax{\rm I\kern-.18em K}}
\def\IP{\relax{\rm I\kern-.18em P}}

\font\cmss=cmss10 \font\cmsss=cmss10 at 7pt
\def\IR{\relax{\rm I\kern-.18em R}}

\def\IZ{\relax\ifmmode\mathchoice
{\hbox{\cmss Z\kern-.4em Z}}{\hbox{\cmss Z\kern-.4em Z}}
{\lower.9pt\hbox{\cmsss Z\kern-.4em Z}}
{\lower1.2pt\hbox{\cmsss Z\kern-.4em Z}}\else{\cmss Z\kern-.4em
Z}\fi}
\def\IB{\relax{\rm I\kern-.18em B}}
\def\IC{{\relax\hbox{$\inbar\kern-.3em{\rm C}$}}}
\def\ID{\relax{\rm I\kern-.18em D}}
\def\IE{\relax{\rm I\kern-.18em E}}
\def\IF{\relax{\rm I\kern-.18em F}}
\def\IG{\relax\hbox{$\inbar\kern-.3em{\rm G}$}}
\def\IGa{\relax\hbox{${\rm I}\kern-.18em\Gamma$}}
\def\IH{\relax{\rm I\kern-.18em H}}
\def\II{\relax{\rm I\kern-.18em I}}
\def\IK{\relax{\rm I\kern-.18em K}}
\def\IP{\relax{\rm I\kern-.18em P}}

\font\cmss=cmss10 \font\cmsss=cmss10 at 7pt
\def\IR{\relax{\rm I\kern-.18em R}}

\def\tilde{\widetilde}
\def\frac#1#2{{{#1} \over {#2}}}

\Title{\vbox{\baselineskip12pt\hbox{hep-th/9707079}
\hbox{IAS-97-74, RU-97-57}}}
{\vbox{\centerline{Matrix Description of Interacting Theories }
\centerline{}
\centerline{in Six Dimensions}}}

\centerline{O. Aharony$^1$, M. Berkooz$^1$, S. Kachru$^2$, 
N. Seiberg$^3$,
and E. Silverstein$^1$}
\smallskip
\smallskip
\centerline{$^1$ Department of Physics and Astronomy}
\centerline{Rutgers University }
\centerline{Piscataway, NJ 08855-0849, USA}
\smallskip
\smallskip
\smallskip
\centerline{$^2$ Department of Physics}
\centerline{University of California at Berkeley}
\centerline{Berkeley, CA 94720, USA}
\smallskip
\smallskip
\smallskip
\centerline{$^3$ Institute for Advanced Study}
\centerline{Princeton, NJ 08540, USA}
\bigskip
\bigskip
\noindent
We propose descriptions of interacting (2,0) supersymmetric 
theories without gravity
in six dimensions in the infinite momentum frame.  They are
based on the large $N$ limit of quantum mechanics or  
1+1 dimensional field theories on the moduli space
of $N$ instantons in $\IR^4$.


\Date{July 1997}

\lref\math{M. Atiyah and J. Jones,
``Topological Aspects of Yang-Mills Theory'',
Comm Math Phys {\bf 61} (1978) 97\semi
C.P. Boyer, J.C. Hurtubise, B.M. Mann and R.J. Milgram,
``The Topology of Instanton Moduli Spaces, I: The Atiyah-Jones
Conjectures'', Ann. Math. {\bf 137} (1993) 561.}

\lref\aps{P.C. Argyres, M.R. Plesser, and N. Seiberg,
``The Moduli Space of Vacua of $N=2$ SUSY QCD and
Duality in $N=1$ SUSY QCD'', hep-th/9603042,
Nucl. Phys. {\bf B471} (1996) 159.}

\lref\newwit{E. Witten, to appear.}

\lref\tori{W. Taylor IV, hep-th/9611042, \pl{394}{1997}{283}.}%
\lref\sus{L. Susskind, hep-th/9611164.}%
\lref\fhrs{W. Fischler, E. Haylo, A. Rajaraman and L. Susskind,
hep-th/9703102.}%
\lref\joe{J. Polchinski, S. Chaudhuri, and C. Johnson, hep-th/9602052; J.
Polchinski, hep-th/9611050.}
\lref\douglas{M.R. Douglas, ``Branes Within Branes'', hep-th/9512077.}
\lref\bss{T. Banks, N. Seiberg and S.H. Shenker, hep-th/9612157,
\np{490}{1997}{91}.}
\lref\ss{S. Sethi and L. Susskind, hep-th/9702101.}%
\lref\grw{O.J. Ganor, S. Ramgoolam and W. Taylor IV, hep-th/9611202.}%
\lref\ees{O. Ganor and A. Hanany,  hep-th/9602120, \np{474}{1996}{122};
N. Seiberg and E. Witten, hep-th/9603003, \np{471}{1996}{121}; M. Duff,
H. Lu and C.N. Pope, hep-th/9603037, \pl{378}{1996}{101}.}

\lref\ktso{S. Govindarajan, hep-th/9705113;  M. Berkooz and M. Rozali,
hep-th/9705175.}

\lref\ADHM{M. Atiyah, V. Drinfeld, N. Hitchin, and Y. Manin,
``Construction of Instantons'',
Phys. Lett. {\bf 65A} (1978) 185.}
\lref\danfer{U. Danielsson and G. Ferretti, ``The Heterotic
Life of the D-particle'', hep-th/9610082.}
\lref\ks{S. Kachru and E. Silverstein, ``On Gauge Bosons in the
Matrix Model Approach to M Theory'', Phys. Lett. {\bf B396} (1997)
70, hep-th/9612162.}

\newsec{Introduction}

\nref\gstring{N. Seiberg, ``New Theories in Six Dimensions and
Matrix Description of M-theory on $T^5$ and $T^5/\IZ_2$'',
hep-th/9705221.}%
\nref\dvv{R. Dijkgraaf, E. Verlinde and H. Verlinde, ``BPS Spectrum of
the Fivebrane and Black Hole Entropy'', hep-th/9603126,
\np{486}{1997}{77}; ``BPS Quantization of the Fivebrane'',
hep-th/9604055, \np{486}{1997}{89}.}%
\nref\juan{J. Maldacena, ``Statistical Entropy of Near Extremal
Five-branes'', Nucl. Phys. {\bf B477} (1996) 168, 
hep-th/9605016.}%
\nref\dvvmarch{R. Dijkgraaf, E. Verlinde and H. Verlinde, ``5D Black
Holes and Matrix Strings'', hep-th/9704018.}%
\nref\brs{M. Berkooz, M. Rozali and N. Seiberg,  ``Matrix Description
of M theory on $T^4$ and $T^5$'', hep-th/9704089.}%

Recent advances in string theory have led to the discovery of many new
theories without gravity.  These theories are found by taking various
limits of the full M theory in which many of the degrees of freedom
decouple.  In particular, by restricting attention to processes with
energy much lower than the Planck scale, $M_p$, we decouple gravity.
In most cases the resulting theory in this limit is simply a free
field theory.  In other cases the low energy theory is an interacting
local quantum field theory at a nontrivial fixed point of the
renormalization group.  Finally, as in \gstring, the limit can yield a
string theory, which is not a standard local quantum field theory
(these theories had been anticipated in \refs{\dvv-\brs}).

The first examples of such nontrivial field theories were found 
\ref\wittentwoz{E. Witten, ``Some Comments on String Dynamics'',
hep-th/9507121, Contributed to STRINGS 95: 
Future Perspectives in String Theory, Los Angeles, CA, 13-18 Mar 
1995.}
in compactifications of the IIB theory on singular K3 surfaces.  These
theories are labeled by the A-D-E singularities of the underlying K3.
The A type theories also appear when several 5-branes of M theory
in eleven dimensions approach each other 
\ref\strominger{A. Strominger, ``Open P-branes'', hep-th/9512059,
\pl{383}{1996}{44}.}, and the D type theories appear when several
5-branes of M theory approach each other at an $\IR^5/\IZ_2$
singularity.

We now review some of the basic properties of these theories (for a
more detailed review see
\ref\sixteen{N. Seiberg, ``Notes on Theories with 16 Supercharges'',
hep-th/9705117.}).
At generic points in the moduli space of these theories 
there are $k$ tensor
multiplets, each consisting of a two form field $B$ with a self-dual
field strength, five real scalars $\Phi_i,i=1,\dots,5$ and a fermion.
The moduli space
of vacua is parametrized by the expectation values of the scalars.  It
is 
\eqn\modspace{{\IR^{5k}\over\CW},}
where $\CW$ is a discrete group.
The theory at the singularities of this space is an interacting local
quantum field theory.  Along the moduli space of vacua the theory has
BPS string like excitations.  Their tensions vanish at the
singularities.  Hence, these theories are often referred to as
tensionless string theories.  However, there is a lot of evidence that
they are standard interacting local quantum field theories.
In particular, these are (2,0) superconformal field theories
with a $Spin(5)$ R-symmetry 
\lref\nahm{W. Nahm, ``Supersymmetries and Their Representations'',
Nucl. Phys. {\bf B135} (1978) 149.} \nahm. 
When these theories are compactified on a
circle of radius $R$, the effective five dimensional theory is a gauge
theory with gauge group $G$ of rank $k$ and $\CW$ is its Weyl group.
The five dimensional gauge coupling is $g_5^2=R$.  For the special
case of $k$ coincident 5-branes the gauge group is $G=U(k)$. 

The five dimensional gauge theory has, in addition to
the gauge bosons, static particle-like BPS
configurations, which are instantons in four dimensions.  Their energy
is $N/R$ where $N$ is the instanton number.  This leads to the
identification of $N$ with the momentum quantum number along the $S^1$
\nref\witteni{E. Witten, ``String Theory Dynamics in
Various Dimensions'', Nucl. Phys. {\bf B443}
(1995) 85, hep-th/9503124.}%
\nref\dkps{M. Douglas, D. Kabat, P. Pouliot, and S. Shenker, 
``D-branes and Short Distances in String Theory'',
Nucl. Phys. {\bf B485} (1997) 85, hep-th/9608024.}%
\nref\rozali{M. Rozali, ``Matrix Theory and U-Duality
in Seven Dimensions'', hep-th/9702136.}%
\refs{\witteni - \rozali}.  As in 
\nref\bfss{T. Banks, W. Fischler, S. Shenker, and L. Susskind, 
``M theory as a Matrix Model:  A Conjecture'', hep-th/9610043,
\physrev{55}{1997}{112}.}%
\nref\lenny{L. Susskind, ``Another Conjecture About Matrix Theory'',
hep-th/9704080.}%
\refs{\bfss} we can use this fact to study the six
dimensional theory in the lightcone frame.  We consider the theory
in the infinite momentum frame and compactify the longitudinal
direction on a circle of radius $R$.  The momentum $N$ along the
circle is the instanton number in the four transverse dimensions.
Let us denote the moduli space of instantons by $\CM_N(G)$.
As in \bfss, we suggest
that the six dimensional (2,0) theory is described by the large $N$
limit of the quantum mechanics on $\CM_N(G)$. The finite $N$ theory
may provide a discrete light cone description of the theory, as in \lenny.

Many people
\ref\manypeople{T. Banks, R. Dijkgraaf,
D. Kutasov, G. Moore, J. Schwarz, S. Sethi, S. Shenker, L. Susskind,
E. Verlinde, H. Verlinde, E. Witten and possibly others, unpublished.}
have independently suggested that the (2,0) field theory and its
stringy extension have a matrix model description. In particular,
there was an inspiring paper \dvvmarch, which is closely related
to our work. We will discuss this further in \S4.

In the following we will describe this quantum
mechanics and some of its generalizations. In section 2 we describe
the quantum mechanical system. In section 3
we will motivate the proposal above by studying the matrix
description of several longitudinal 5-branes along the lines of
\ref\bd{M. Berkooz and M. Douglas, ``Five-branes in M(atrix) Theory'',
hep-th/9610236.}.
We will also motivate it by examining the matrix description of IIB
compactifications on singular K3 surfaces along the lines of
\gstring\ or its low energy limit
\ref\spkthr{A. Fayyazudin and D.J. Smith, 
``A Note on $T^5/\IZ_2$ Compactification 
of the M-theory Matrix Model'', hep-th/9703208; N. Kim and
S.-J. Rey, ``M(atrix) Theory on $T^5/\IZ_2$ Orbifold and
Fivebrane'', hep-th/9705132.}.
In both cases we can scale out gravity to derive the prescription
above. Section 4 contains generalizations of this
construction to the string theories of \gstring.

\newsec{The Quantum Mechanical System}

Since we are going to describe a space time theory with 16
supercharges we should study a quantum mechanical system with 8
supersymmetries.  The maximal R symmetry is $Spin(8)$ but our system
will have only a $Spin(3)\times Spin(5)$ subgroup under which the
supercharges transform as $({\bf 2}, {\bf 4})$.  One way to
construct such theories is to start with a (1,0) supersymmetric field
theory in six dimensions and to dimensionally reduce it to quantum
mechanics.  The first $Spin(3) \cong SU(2)$ factor is an R symmetry,
which is present already in six dimensions.  The $Spin(5)$ factor
appears from the dimensional reduction.  (These theories were recently
discussed in
\ref\andia{D. Diaconescu and R. Entin, ``A Non-Renormalization
Theorem for the d=1, N=8 Vector Multiplet'', hep-th/9706059.}.)

For describing the theories with $G=U(k)$ we use a $U(N)$ gauge theory
with hypermultiplets in the adjoint representation and $k$
fundamentals.  This theory has an $SU(k)\times SU(2) \times SU(2)
\times Spin(5)$ global symmetry, where the first $SU(2)$ factor acts
on the field in the adjoint representation and the second is part of
the R symmetry.  For the theories with $G=SO(2k)$ we use an $Sp(N)$
gauge theory with hypermultiplets in an antisymmetric tensor and $k$
fundamentals.  This theory has an $SO(2k)\times SU(2) \times SU(2)
\times Spin(5)$ global symmetry, where the first $SU(2)$ factor acts on
the field in the antisymmetric representation.  These theories have
Higgs branches, in which both types of hypermultiplets obtain VEVs,
which are isomorphic to the moduli spaces $\CM_N(SU(k))$ and
$\CM_N(SO(2k))$, respectively \ADHM.  The $Spin(5)$ factor in the
global symmetry does not act on the Higgs branch (it acts only on the
fermions).  The other global symmetries are spontaneously broken to a
subgroup at a generic point in the moduli space.

The quantum mechanical systems, which we are interested in, differ
from these in two ways.  First, along the Higgs branch the gauge
theories have effects from the massive gauge bosons, which have been
integrated out.  These are not present in the minimal quantum
mechanical system with target space $\CM_N(G)$.  Second, the gauge
theories have Coulomb branches emanating from the singularities of
$\CM_N(G)$.  These two differences are removed, if we consider the
limit that the gauge coupling $g_{QM}$ goes to infinity.  This gives
the Higgsed gauge multiplets infinite masses and, therefore, decouples
the Coulomb branch from the interior of the Higgs branch.

The $g_{QM}\rightarrow \infty$ limit is the IR limit of the quantum
mechanics.  In this limit all the higher derivative operators in the
effective action are negligible.  Furthermore, as in \aps, we can
promote $1/g_{QM}^2$ to a background vector superfield and use the
decoupling of vector superfields from the Higgs branch to show that
the metric on the Higgs branch is independent of $g_{QM}$.  This
nonrenormalization theorem shows that the strong coupling limit of the
quantum mechanics is simply the supersymmetric quantum mechanics with
target space $\CM_N(G)$.
It should be stressed that unlike higher dimensional field theories,
in quantum mechanics the singularities in $\CM_N(G)$ are quite tame.
They merely limit the wave functions to be single valued around them.

We can generalize this construction for other
gauge groups $G$ by studying the supersymmetric
quantum mechanics
with target space $\CM_N(G)$.  The global symmetry of this theory is
$G \times SU(2) \times SU(2) \times Spin(5)$, where as before the
$Spin(5)$ symmetry acts only on the fermions.

\newsec{Motivation From M Theory}

We will now motivate this proposal by studying $k$ coincident 5-branes
in eleven dimensions.  The theory along the 5-branes is the (2,0)
theory of $U(k)$.  Let us compactify the longitudinal direction along
the 5-branes on a circle of radius $R$.  The $k$ longitudinal branes
are now D4-branes.  The system with $N$ units of momentum around the
circle is then described by $k$ D4-branes and $N$ D0-branes.  This
system was studied in detail in \dkps.  The D0-branes are described by
a quantum mechanical system with a $U(N)$ gauge symmetry with
hypermultiplets in the adjoint and $k$ fundamentals.  The Coulomb
branch corresponds to the motion of the D0-branes away from the
D4-branes.  In matrix theory, graviton states live on this
branch.  The Higgs branch corresponds to D0-branes in the D4-branes
where they can grow.  This fact can be interpreted by a D4-brane
observer as instantons in the $U(k)$ gauge theory on the D4-brane 
\douglas.

This quantum mechanical system was used in \bd\ to describe the
longitudinal 5-brane in the matrix theory.  Since these theories can
be viewed as dimensional reduction of a higher dimensional gauge
theory, we find it convenient to
rescale all the scalars to have dimension one.  Then, the gauge
coupling of the quantum mechanical system is
\eqn\gqm{g_{QM}^2 = M_p^6 R^3.}

The authors of \bd\ studied the case $k=1$, where there is no Higgs
branch in the quantum mechanical system\foot{There is always an
$\IR^4$ component of the Higgs branch which corresponds to the center
of mass of all the instantons inside the 5-brane. This arises in the
gauge theory from the singlet component of the adjoint hypermultiplet
(for $U(N)$) or of the antisymmetric multiplet (for $Sp(N)$), and it
will give rise to the spacetime on the 5-brane in our description.}.  The
Coulomb branch was interpreted there as leading to the
space time transverse to the fivebrane.  We are
interested in larger values of $k$, where there is also a Higgs
branch.  Should it be interpreted as leading to a new branch of space
time?  In order to answer this question we should recall how space
time appears in the quantum mechanical systems
\nref\bound{E. Witten, 
``Bound States of Strings and P-branes'',
\np{460}{1996}{335}, hep-th/9510135.}%
\refs{\bound,\bfss}.  One studies low energy states in the quantum
system, which are localized far from the singularities and interprets
them as particles moving in space time.  The noncompact nature of the
Higgs branch might lead to an interpretation as a new branch of space
time.  We would like to suggest, however, a different interpretation.
The degrees of freedom along the Higgs branch correspond to states of
the interacting (2,0) theory (other states may be concentrated at the
singularities).  As an interacting conformal field theory, this theory
does not have a particle interpretation.  Therefore, we should not
interpret the Higgs branch as a new branch of space time.  As the
space time moduli are changed so that two 5-branes coincide, the Higgs
branch opens up.  This leads to IR divergences in the S matrix
corresponding to the lack of well defined particle states and S
matrix.  It will be interesting to understand explicitly how the
quantum mechanics along the Higgs branch captures the dynamics of the
(2,0) field theory, and how to extract the operators and correlation
functions of this theory.

In order to focus on the degrees of freedom of the (2,0) field theory
we should consider the limit $M_p \rightarrow \infty$ holding all
other energies fixed.  In this limit the degrees of freedom on the
eleven dimensional five brane decouple from the degrees of freedom in
the bulk of space time.  In particular, they decouple from gravity.
Now, it is clear from \gqm\ that in this limit $g_{QM}$ goes to
infinity.  As we discussed above, this limit removes the Coulomb
branch and restricts the quantum mechanics to the Higgs branch\foot{In
the case $G=U(k)$ corresponding to $k$ coincident fivebranes, there is
also a decoupled free tensor multiplet, which has asymptotic
scattering states.  The corresponding states in the quantum mechanics
come from the singularities at the intersection of the Higgs and
Coulomb branches.  These also decouple from the interacting SCFT.}.
The $Spin(5)$ R-symmetry in the superconformal algebra arises from the
$Spin(5)$ R-symmetry in the quantum mechanics discussed in \S2.  The
$SU(2) \times SU(2)$ global symmetry of the quantum system is
interpreted as the transverse Lorentz symmetry.  It is spontaneously
broken on the Higgs branch.  This corresponds to the ``breaking'' of
Lorentz invariance by the instanton positions in the transverse
spacetime.

We can also derive the same conclusion by examining the
compactification of the type IIB theory on a singular K3, which is dual
to M theory on $T^5/\IZ_2$
\ref\tbkt{K. Dasgupta and S. Mukhi, ``Orbifolds
of M-theory'', hep-th/9512196,
\np{465}{1996}{399}; E. Witten, ``Five-branes and M-theory on an
Orbifold'', hep-th/9512219, \np{463}{1996}{383}.}.  A matrix model
description of this theory \gstring\ is based on
heterotic fivebranes at zero string coupling,
wrapped on a $T^5$ with cycles of length 
$\Sigma_1, \dots, \Sigma_5$.  In a region of the moduli space,
at low energies,
this is an $Sp(N)$ gauge theory in six dimensions, with 
16 fundamental hypermultiplets and one hypermultiplet
in the antisymmetric tensor 
\ref\sminst{E. Witten, ``Small Instantons in String Theory'',
hep-th/9511030, \np{460}{1995}{541}.}.  
The gauge coupling of this theory is given
by $1/g_6^2 = M_s^2$.  At energies
of the order of the heterotic string scale $M_s$, string degrees
of freedom become important.

The masses of the fundamentals correspond to some of the moduli of the
space time theory. In the interpretation of the theory as M theory on
$T^5/\IZ_2$, they are related to the positions of the 16 5-branes in
spacetime \tbkt. When $k$ of these masses are equal, the K3 has an
$A_{k-1}$ singularity and we find the (2,0) conformal theory
associated with $U(k)$.  The gauge theory near this singularity is the
$U(N)$ theory described in the previous section. When $k$ of the
masses are zero, the K3 has a $D_k$ singularity leading to the
$SO(2k)$ (2,0) theory.  These theories are only nontrivial for $k>1$.
Then, there are finite-size instantons in the corresponding Yang-Mills
theory, and in the quantum mechanics there is a Higgs branch
isomorphic to the moduli space of these instantons.  In the context of
the full six dimensional string theory of \gstring\ this can be
extended to exceptional groups corresponding to $E_{6,7,8}$
singularities of the underlying K3.

In the spacetime theory, we would like to scale out
gravity, i.e. send $M_p \rightarrow \infty$.
We can now consult the formulas relating the spacetime quantities
to the matrix theory parameters \gstring\ to determine
the corresponding limit in the matrix theory.
Taking a square $T^5/\IZ_2$ with lengths $L_1,\dots,L_5$ in
spacetime, and letting the radius of the longitudinal
direction be $R$, the map is
\eqn\mapI{\Sigma_i={1\over{M_p^3RL_i}}}
\eqn\mapII{M_s^2=M_p^9R^2L_1L_2L_3L_4L_5.}
Now let us take $M_p\rightarrow \infty$.  Then, $M_s\to\infty$, so the
string degrees of freedom in the matrix theory decouple.
At the same time, $\Sigma_i\to 0$, so the matrix theory
reduces to quantum mechanics.  The gauge coupling $g_{QM}$ in
this quantum mechanics behaves as
\eqn\couplim{{1 \over g_{QM}^2}=
M_s^2\Sigma_1\Sigma_2\Sigma_3\Sigma_4\Sigma_5 =
{1\over M_p^6R^3}\to 0.}

So we have $g_{QM}\to\infty$, which means, as discussed above,
that the Coulomb branch decouples from the Higgs branch.
We see then that in the limit in which spacetime gravity decouples
from the theory on the fivebranes, the Coulomb branch in the matrix
quantum mechanics decouples from the Higgs branch.  Thus,
the quantum field theory on the fivebrane is reformulated 
a la matrix theory as
the quantum mechanics on the moduli space of instantons. Note that
deriving the $E_n$ theories in this way is more complicated, since in
the M theory description they require taking some of the $L_i$ to be
small, but we still expect the description to go over to the quantum
mechanics on the moduli space of instantons as described in the
previous section.

\newsec{A Description of the Six
Dimensional String Theories}

A simple generalization of the methods used in the previous sections
provides a description\foot{Similar ideas have been independently
studied by E. Witten \newwit.} of the new string theories in six
dimensions
\gstring.  To derive these theories from M theory, we look at $k$
5-branes on a transverse circle of radius $L_1 \to 0$.  We take a
scaling limit such that $\tilde{M}_{string}^2 = L_1 M_p^3$ (which is
the tension of membranes wrapped around this circle) remains constant
\gstring. As in the previous section, we can describe fivebranes in M
theory by going to particular points in the moduli space of M theory
on $T^5/\IZ_2$, where $k$ fivebranes come together.

Starting with the matrix theory description of this in terms of
$Spin(32) / \IZ_2$ heterotic fivebranes on $T^5$, we can determine the
limit corresponding to this string theory by using the relations
\mapI\ and \mapII, and get a description of
this string theory. We find that we need to
take the size of the circles $\Sigma_{2,3,4,5}$ to
zero, take $M_s^2$ to infinity, and scale $\Sigma_1$ as
$1 / R \tilde{M}_{string}^2$. The resulting
1+1 dimensional gauge coupling is
given by 
\eqn\coupling{{1\over g_{2}^2} = M_s^2
\Sigma_2 \Sigma_3 \Sigma_4 \Sigma_5 = {L_1 \over R^2 M_p^3},}
so again we find that the description involves the $g_{2} \to \infty$
limit of the gauge theory. In this case, however, $\Sigma_1$ is kept
finite, so that the description is in terms of a $1+1$ dimensional
gauge theory.  The $g_2 \rightarrow \infty$ limit is the IR limit of
this theory.  In this limit the theory becomes scale invariant.
However, the presence of the finite circle $\Sigma_1$ creates a scale.
Equivalently, we can set it to one and measure all other scales
relative to it.

Note that we are taking
the 5-brane to be far from the orbifold points, so the gauge theory is
simply a $U(N)$ gauge theory with an adjoint hypermultiplet and $k$
fundamental hypermultiplets, as in section 2. When we are far from the
singularities the fact that we started from M theory on $T^5/\IZ_2$
and not in flat space should be unimportant, so we expect to get the
usual string theory corresponding to a type IIA 5-brane in spacetime.

The theory we have found in the scaling limit basically recovers the
theory discussed in \dvvmarch.  These authors studied the theory on
the fivebrane in terms of a matrix string
theory 
\nref\motl{L. Motl, ``Proposals on Nonperturbative Superstring 
Interactions'', hep-th/9701025.}%
\nref\bs{T. Banks and N. Seiberg, ``Strings from Matrices'',
hep-th/9702187.}%
\nref\dvvmatrix{R. Dijkgraaf, E. Verlinde and H. Verlinde, ``Matrix
String Theory'', hep-th/9703030.}%
\refs{\motl-\dvvmatrix}.  The main new element that we add
is the limit $g_2 \rightarrow \infty$.  This limit can be motivated as
we did above.  Alternatively, we are trying to study the string
theories of \gstring, where the string coupling was taken to zero.  In
terms of the matrix string theory this is precisely the limit where
$g_2 \rightarrow \infty$.  This limit enabled us to decouple the Higgs
branch from the Coulomb branch.  This is analogous to the statement in
\gstring\ that in this limit the fundamental strings are trapped in
the fivebrane (the Coulomb branch corresponds to motion of the strings
away from the fivebrane).

There are a few other differences between our proposal and that of
\dvvmarch.  We consider moduli spaces of instantons on $\IR^4$,
instead of instantons on $T^4$.  More importantly, as we explained
above, the nonrenormalization theorem of \aps\ guarantees that the
Higgs branch of our quantum field theory receives no quantum
corrections.  Therefore there is no room for any marginal operator to
correct $\CM_N(G)$.

Other novel six dimensional theories (without gravity)
can be defined in a similar
fashion. For instance, the theory of a 5-brane in M theory on a
transverse $T^2$
with $M_p \to \infty$ and $L_{1,2} M_p^3$ kept constant would be
described by a $2+1$ dimensional field theory compactified on two
circles whose length scales as $1/R$. 

Unfortunately, we can only
describe in this way the six dimensional theories and not their
compactifications, which are useful for Matrix theory descriptions of
M theory compactifications \refs{\brs,\gstring}. Since these theories
are non-local, their definition in infinite six dimensional space does
not completely describe their compactifications, so that a concrete
description of the compactified theories is still an interesting open
problem.

\bigskip

\centerline{\bf Acknowledgments}\nobreak

We would like to thank P. Aspinwall, T. Banks, M. Douglas, D. Freed,  
K. Intriligator, J. Milgram, G. Moore, S. Sethi, S. Shenker, H. Verlinde, 
and E. Witten for
useful discussions.  This work was supported in part by DOE grants
\#DE-FG02-90ER40542 and \#DE-FG02-96ER40559.

\listrefs
\end